\documentclass[oneside]{cernyrep}

\pdfoutput=1 

\usepackage[colorinlistoftodos]{todonotes}
\usepackage{ifthen}  
\usepackage{hyperref}
\newboolean{uprightparticles} 
\setboolean{uprightparticles}{false} 

\usepackage{xspace} 
\usepackage{upgreek}

\def\MagUp {\mbox{\em Mag\kern -0.05em Up}\xspace}

\ifthenelse{\boolean{uprightparticles}}%
{

 \def\PDelta      {\ensuremath{\Delta}\xspace}                 
 \def\PXi      {\ensuremath{\Xi}\xspace}                 
 \def\PLambda      {\ensuremath{\Lambda}\xspace}                 
 \def\PSigma      {\ensuremath{\Sigma}\xspace}                 
 \def\POmega      {\ensuremath{\Omega}\xspace}                 
 \def\PUpsilon      {\ensuremath{\Upsilon}\xspace}                 
 
 \def\PB      {\ensuremath{\mathrm{B}}\xspace}                 
                  
 \def\PD      {\ensuremath{\mathrm{D}}\xspace}

 \def\PH      {\ensuremath{\mathrm{H}}\xspace}

 \def\PK      {\ensuremath{\mathrm{K}}\xspace}

 \def\PN      {\ensuremath{\mathrm{N}}\xspace}

 \def\PW      {\ensuremath{\mathrm{W}}\xspace}                 
 \def\PX      {\ensuremath{\mathrm{X}}\xspace}                 
                  
 \def\PZ      {\ensuremath{\mathrm{Z}}\xspace}                 
 \def\Pa      {\ensuremath{\mathrm{a}}\xspace}                 
 \def\Pb      {\ensuremath{\mathrm{b}}\xspace}

 \def\Pe      {\ensuremath{\mathrm{e}}\xspace}                 
 \def\Pf      {\ensuremath{\mathrm{f}}\xspace}                 
 \def\Pg      {\ensuremath{\mathrm{g}}\xspace}                 
 \def\Ph      {\ensuremath{\mathrm{h}}\xspace}                 
 \def\Pi      {\ensuremath{\mathrm{i}}\xspace}

 \def\Pl      {\ensuremath{\mathrm{l}}\xspace}

}
{

 \mathchardef\PDelta="7101
 \mathchardef\PXi="7104
 \mathchardef\PLambda="7103
 \mathchardef\PSigma="7106
 \mathchardef\POmega="710A
 \mathchardef\PUpsilon="7107
                  
 \def\PB      {\ensuremath{B}\xspace}                 
                  
 \def\PD      {\ensuremath{D}\xspace}

 \def\PH      {\ensuremath{H}\xspace}

 \def\PK      {\ensuremath{K}\xspace}

 \def\PN      {\ensuremath{N}\xspace}

 \def\PW      {\ensuremath{W}\xspace}                 
 \def\PX      {\ensuremath{X}\xspace}                 
                  
 \def\PZ      {\ensuremath{Z}\xspace}                 
 \def\Pa      {\ensuremath{a}\xspace}                 
 \def\Pb      {\ensuremath{b}\xspace}

 \def\Pe      {\ensuremath{e}\xspace}                 
 \def\Pf      {\ensuremath{f}\xspace}                 
 \def\Pg      {\ensuremath{g}\xspace}                 
 \def\Ph      {\ensuremath{h}\xspace}                 
 \def\Pi      {\ensuremath{i}\xspace}

 \def\Pl      {\ensuremath{l}\xspace}

}

\makeatletter
\ifcase \@ptsize \relax
  \newcommand{\miniscule}{\@setfontsize\miniscule{4}{5}}
\or
  \newcommand{\miniscule}{\@setfontsize\miniscule{5}{6}}
\or
  \newcommand{\miniscule}{\@setfontsize\miniscule{5}{6}}
\fi
\makeatother

\DeclareRobustCommand{\optbar}[1]{\shortstack{{\miniscule (\rule[.5ex]{1.25em}{.18mm})}
  \\ [-.7ex] $#1$}}

  \def\Kbar    {{\kern 0.2em\overline{\kern -0.2em \PK}{}}\xspace}

\def\KorKbar    {\kern 0.18em\optbar{\kern -0.18em K}{}\xspace}

  \def\Dbar    {{\kern 0.2em\overline{\kern -0.2em \PD}{}}\xspace}

\def\DorDbar    {\kern 0.18em\optbar{\kern -0.18em D}{}\xspace}

\def\Bbar    {{\ensuremath{\kern 0.18em\overline{\kern -0.18em \PB}{}}}\xspace}

\def\BorBbar    {\kern 0.18em\optbar{\kern -0.18em B}{}\xspace}

  \def\Y#1S{\ensuremath{\PUpsilon{(#1S)}}\xspace}

\def\Lbar        {{\ensuremath{\kern 0.1em\overline{\kern -0.1em\PLambda}}}\xspace}
\def\LorLbar    {\kern 0.18em\optbar{\kern -0.18em \PLambda}{}\xspace}

\def\to                 {\ensuremath{\rightarrow}\xspace}

\newcommand{\as}{{\ensuremath{\alpha_s}}\xspace}

\def\AT#1     {\ensuremath{A_{\mathrm{T}}^{#1}}\xspace}           

\def\C#1      {\ensuremath{\mathcal{C}_{#1}}\xspace}                       
\def\Cp#1     {\ensuremath{\mathcal{C}_{#1}^{'}}\xspace}                    
\def\Ceff#1   {\ensuremath{\mathcal{C}_{#1}^{\mathrm{(eff)}}}\xspace}        
\def\Cpeff#1  {\ensuremath{\mathcal{C}_{#1}^{'\mathrm{(eff)}}}\xspace}       
\def\Ope#1    {\ensuremath{\mathcal{O}_{#1}}\xspace}                       
\def\Opep#1   {\ensuremath{\mathcal{O}_{#1}^{'}}\xspace}                    

\newcommand{\tev}{\ifthenelse{\boolean{inbibliography}}{\ensuremath{~T\kern -0.05em eV}\xspace}{\ensuremath{\mathrm{\,Te\kern -0.1em V}}}\xspace}
\newcommand{\gev}{\ensuremath{\mathrm{\,Ge\kern -0.1em V}}\xspace}
\newcommand{\mev}{\ensuremath{\mathrm{\,Me\kern -0.1em V}}\xspace}
\newcommand{\kev}{\ensuremath{\mathrm{\,ke\kern -0.1em V}}\xspace}
\newcommand{\ev}{\ensuremath{\mathrm{\,e\kern -0.1em V}}\xspace}
\newcommand{\gevc}{\ensuremath{{\mathrm{\,Ge\kern -0.1em V\!/}c}}\xspace}
\newcommand{\mevc}{\ensuremath{{\mathrm{\,Me\kern -0.1em V\!/}c}}\xspace}
\newcommand{\gevcc}{\ensuremath{{\mathrm{\,Ge\kern -0.1em V\!/}c^2}}\xspace}
\newcommand{\gevgevcccc}{\ensuremath{{\mathrm{\,Ge\kern -0.1em V^2\!/}c^4}}\xspace}
\newcommand{\mevcc}{\ensuremath{{\mathrm{\,Me\kern -0.1em V\!/}c^2}}\xspace}
\newcommand{\kevcc}{\ensuremath{{\mathrm{\,ke\kern -0.1em V\!/}c^2}}\xspace}

\def\cm   {\ensuremath{\mathrm{ \,cm}}\xspace}

\def\mm   {\ensuremath{\mathrm{ \,mm}}\xspace}

\def\nm   {\ensuremath{\mathrm{ \,nm}}\xspace}

\def\barn{\ensuremath{\mathrm{ \,b}}\xspace}

\def\invnb {\ensuremath{\mbox{\,nb}^{-1}}\xspace}

\def\invpb {\ensuremath{\mbox{\,pb}^{-1}}\xspace}

\def\invfb   {\ensuremath{\mbox{\,fb}^{-1}}\xspace}

\def\invab   {\ensuremath{\mbox{\,ab}^{-1}}\xspace}

\def\ns   {\ensuremath{{\mathrm{ \,ns}}}\xspace}
\def\ps   {\ensuremath{{\mathrm{ \,ps}}}\xspace}
\def\fs   {\ensuremath{\mathrm{ \,fs}}\xspace}

\def\gsim{{~\raise.15em\hbox{$>$}\kern-.85em
          \lower.35em\hbox{$\sim$}~}\xspace}
\def\lsim{{~\raise.15em\hbox{$<$}\kern-.85em
          \lower.35em\hbox{$\sim$}~}\xspace}

\def\mrad{\ensuremath{\mathrm{ \,mrad}}\xspace}
\def\rad{\ensuremath{\mathrm{ \,rad}}\xspace}

\def\gauss      {\mbox{\textsc{Gauss}}\xspace}

\def\tell1  {TELL1\xspace}
\def\ukl1   {UKL1\xspace}

\title{\vspace*{0.1cm}\LARGE SHiP experiment at the SPS Beam Dump Facility}

\author{\includegraphics[height=20mm]{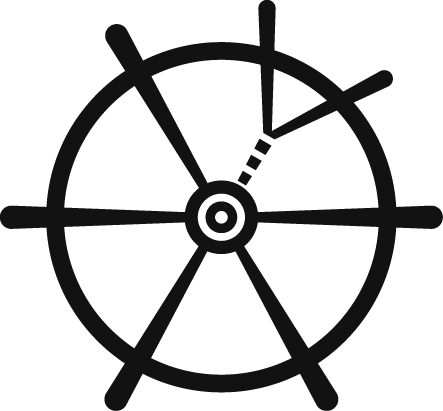} SHiP Collaboration \& HI-ECN3 Project Team}

\begin{abstract}

\end{abstract}

\begin{document}
\pagestyle{plain}
\pagenumbering{arabic}

\clubpenalty = 10000
\widowpenalty = 10000
\displaywidowpenalty = 10000

\thispagestyle{empty}
\setlength{\unitlength}{1mm}
\begin{picture}(0.001,0.001)
\put(120,-14){\includegraphics[height=30mm]{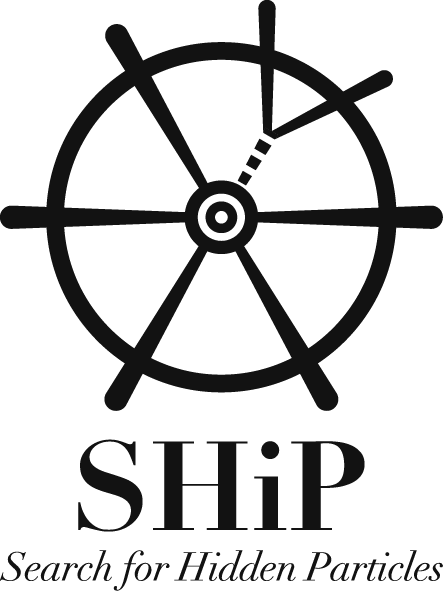}}



\put(8,-38){\LARGE\bfseries
              SHiP experiment at the SPS Beam Dump Facility 
              }
              
\put(22,-57){\Large
              $^1$SHiP Collaboration  \& HI-ECN3 Project Team
              }


\put(59,-86){
                {\large\bf Abstract}
             }

\put(2,-96){
\parbox[t][30mm]{140mm}{
In 2024, the SHiP experiment, together with the associated Beam Dump Facility (BDF) under the auspices of the High Intensity ECN3 (HI-ECN3) project, was selected for the future physics exploitation of the ECN3 experimental facility at the SPS. The SHiP experiment is a general-purpose intensity-frontier setup designed to search for physics beyond the Standard Model in the domain of Feebly Interacting Particles at the GeV-scale.  
It comprises a multi-system apparatus that provides discovery sensitivity to both decay and scattering signatures of models with feebly interacting particles, such as dark-sector mediators, both elastic and inelastic light dark matter, as well as millicharged particles. The experiment will also be able to perform both Standard Model measurements and Beyond Standard Model searches with neutrino interactions. In particular, it will have access to unprecedented statistics of tau and anti-tau neutrinos. The construction plan foresees commissioning of the facility and detector, and start of operation in advance of Long Shutdown 4, with a programme of exploration for 15 years of data taking. By exploring unique regions of parameter space for feebly interacting particles in the GeV/c$^2$ mass range, the SHiP experiment will complement ongoing searches at the LHC and searches at future colliders.

}}

\put(3,-175){\small Document submitted to European Strategy for Particle Physics Update 2026}

\put(0,-230){\small Contacts: Andrei.Golutvin@cern.ch, Matthew.Fraser@cern.ch, Richard.Jacobsson@cern.ch}

\put(0,-240){\small $^1$Complete author lists at the end}
\end{picture}

\newpage
\vfill  \normalsize


\setcounter{page}{1}

\section{Introduction}

The discovery of the Higgs boson at the LHC in 2012 
marked the completion of the Standard Model (SM) of particle physics. Since then, extensive experimental efforts have tested the properties of the Higgs boson, and so far, all results remain consistent with SM predictions. This success is mirrored in cosmology, where the $\Lambda$-Cold Dark Matter model, based on Einstein's General Relativity, describes the large-scale structure of the Universe with remarkable precision. However, while the SM provides an accurate description of known elementary particles and their interactions, it does not account for several key phenomena observed in nature. Several fundamental questions—such as the nature of dark matter, the origin of neutrino masses, and the matter-antimatter asymmetry of the Universe—demand the existence of physics beyond the Standard Model (BSM). The lack of any evidence or hints for new physics so far suggests two possible scenarios: either new physics resides at energy scales beyond current accelerator reach, or it is only endowed with extremely feeble interactions that have so far remained undetectable.

{\bf The need for dedicated searches for feebly-interacting particles (FIPs) and long-lived new physics states at accelerators was explicitly recognized by the 2020 update of the European Strategy for Particle Physics (ESPP)}. It emphasized the importance of exploring the hidden sector as part of a broad and balanced experimental programme. Following this recommendation, CERN examined multiple proposals for intensity-frontier experiments aimed at probing feebly-coupled new physics, culminating in the {\bf selection of the SHiP experiment~\cite{Bonivento:2013jag, Anelli:2015pba, Alekhin:2015byh, SHiP_CDS, SHiP_detector, SHiP_ECN3}(and references therein) and the approval of the HI-ECN3 project in June 2024, with ratification by the CERN Council}. SHiP now stands as a key element of CERN’s strategy to expand beyond the energy frontier, complementing ongoing and future high-intensity and precision-frontier searches.

The SHiP experiment is specifically designed to explore feebly-interacting long-lived particles, capable of covering regions inaccessible at high-energy colliders. By targeting a broad class of FIPs — including heavy neutral leptons (HNLs), dark photons, dark scalars, axion-like particles (ALPs), and light dark matter (LDM) — SHiP provides a dedicated facility on the "Coupling Frontier" to systematically probe a vast, unexplored parameter space of new long-lived particles.

The experiment will be hosted at a  new SPS Beam Dump Facility (BDF)~\cite{SHiP:2018yqc, BDF_YELLOWBOOK, LLOstudy} to be constructed in the upgraded ECN3 experimental facility, leveraging the high-intensity SPS proton beam to maximise the physics reach. SHiP  will exploit the potential for hidden-sector particle production directly through inelastic processes, in electromagnetic processes, and decays of light and heavy mesons generated in the beam dump. These particles, which would typically escape detection at colliders, can be studied at SHIP in a controlled environment where their decays or interactions occur in a shielded, low-background detector.

Beyond its role in hidden-sector physics, SHiP will provide the world’s most intense neutrino flux in the multi-GeV range, enabling unique SM tests and BSM searches with neutrinos. The experiment will collect an unprecedented sample of tau and anti-tau neutrino interactions, allowing for high-precision measurements.


With the approval of the Technical Design Report (TDR) phase, SHiP is now positioned to transition towards implementation. Over the last year, the collaboration and the HI-ECN3 project have made significant progress in key technical milestones, including detector prototyping, as well as beamline and target-system design,  ensuring timely readiness for construction and commissioning. The experimental timeline is aligned with CERN’s accelerator schedule. The upgrade of ECN3 will start in Long Shutdown 3 with the dismantling of the existing experimental setup before the installation of BDF/SHiP with commissioning and first data-taking planned for 2031-2033, well before Long Shutdown 4 (LS4), currently scheduled for 2034. {\bf The initial data-taking phase in 2032-2033 is critical to validate background conditions and detector performance, and to provide the first physics results with significant discovery potential well beyond current experimental bounds in time for the final decisions on the experiments for the future colliders}. 

The full physics programme of SHiP will then be able to exploit the facility for 15 years of data-taking, establishing SHiP as a cornerstone of CERN’s long-term strategy in non-energy-frontier particle physics. {\bf By extending CERN’s reach into the coupling frontier, SHiP strengthens Europe’s leadership in intensity-frontier research,  
ensuring that all viable paths to new physics remain rigorously explored}.


\section{Physics programme at the "Coupling Frontier"}
\label{sec:physicscase}

The parameter space for potential new particles that could extend the SM spans a vast range of masses and coupling constants, requiring a broad-gauged experimental approach.
The SHiP experiment is designed to explore this uncharted territory by pushing the limits of the coupling frontier. It will search for signatures of FIPs, including LDM, and perform precision measurements in neutrino physics, leveraging the high-intensity proton beam at the SPS. 

The main BSM target of SHiP are FIPs with mass in the 0.5\,--\,5\,GeV/c$^2$ range. 
This domain is special and attractive for several reasons: (i) current collider experiments and experiments at future colliders have limited potential to explore this range; (ii) the possibility to copiously produce these particles by decays of numerous mesons and have thousands of events, which may allow, in case of discovery, to reconstruct the properties of the FIP and test if it is involved in the resolution of the BSM problems; (iii) synergy between laboratory probes and cosmological/astrophysical observations, that define the target FIP parameter space in terms of coupling from below and from above.

SHiP's search for FIP decay signatures is designed to be performed in a near-zero background environment, while the measurements of scattering signatures have a very low but well-controlled irreducible background component from neutrino interactions. These are crucial conditions given that the expected event rates for many of these feebly-interacting particles are extremely low.

The following sections outline SHiP's physics objectives and experimental strategy in more detail.

\subsection{Searches for FIPs through decay signatures}

The discovery of long-lived FIPs would fundamentally transform our understanding and strategy in particle physics. These particles, predicted in many well-motivated extensions of the SM, could provide crucial insights into some of the most profound open questions, such as the origin of neutrino masses, the nature of dark matter, and the mechanism behind baryogenesis~\cite{Asaka:2005an}.

With the use of the SPS high-intensity proton beam and a detector capable of reconstructing a very wide range of multi-body final states, SHiP is specifically designed to directly and generically search for decays of FIPs that couple feebly to SM particles. Some of the most compelling candidates include (see~\cite{Alekhin:2015byh} and references therein):
\begin{itemize}
    \item {\bf Heavy Neutral Leptons (HNLs)}: Predicted in seesaw models, HNLs offer a natural explanation for neutrino oscillations and masses and the observed matter-antimatter asymmetry via leptogenesis;
    \item {\bf Dark Scalars and Dark Photons}: Emerging from Higgs-portal and vector-portal models, respectively, these particles provide a potential link between the SM and a hidden sector, could play a role in explaining dark matter, or be involved in inflation;
    \item {\bf Axion-like Particles (ALPs)}: ALPs arise in many extensions of the SM, and can be viable dark matter candidates, or mediate the interactions between SM and hidden sectors;
    \item[--] {\bf Inelastic dark matter}: generically appear when considering the scenario with dark matter escaping direct detection; 
    \item[--] {\bf SUSY}: depending on the super-symmetry breaking scale, super-partners like neutralinos may have mass in the GeV/c$^2$ range.
\end{itemize}

The observation of a GeV/c$^2$-mass FIP may also shed light on the existence of new physics at much higher scales. For instance, ALPs couple to the SM via non-renormalizable operators, with the dimensional coupling closely related to the scale of the ultraviolet completion of the model. Detecting ALPs and extracting the value of the coupling would allow identifying the scale, similar to how meson factories constrain new physics scales by the non-observation of modifications to rare meson decays.


To achieve the highest sensitivity to new physics, SHiP implements a multilayered background suppression and control strategy, combining optimized shielding, high-granularity background tagger systems, precise timing and tracking information, and particle identification. One of the primary sources of background originates from neutrino-induced interactions, where high-energy neutrinos produced in the proton-beam interactions may generate secondary particles that mimic FIP decays. Another significant background source is the high flux of muons generated in the beam dump. They are capable of producing background through deep-inelastic interactions with the material of the detector, but also combinatorial background from random muons forming fake vertices.   
Cosmic-ray-induced backgrounds are effectively mitigated by the underground location of SHiP and the surrounding background tagger systems, making their contribution negligibly small.

The SHiP experiment uses a heavy and dense target and a hadron absorber to minimise muons and neutrinos coming from decay in flight of pions and kaons, and a dedicated muon shield that significantly reduces the flux of muons before they reach the fiducial volume of the detector systems. These are complemented by high-efficiency background tagger systems, capable of rejecting interactions that could produce background-like signals. A large helium-filled decay region further minimizes material interactions that could contribute to secondary backgrounds in the search for decay signatures, while precise timing and topological selection criteria provide additional discrimination power.

The SHiP experiment is uniquely positioned to explore vast regions of parameter space that remain unconstrained. Its ability to detect long-lived particles will extend well below the coupling strengths currently probed by accelerator-based experiments. In the benchmarks tested, SHiP's sensitivity reaches 2-4 orders of magnitude in the coupling scale beyond current experimental bounds. For fixed couplings, the span in mass may be over 1-2 orders of magnitude. Hence, SHiP can measure thousands of events for couplings that are orders of magnitude below the current limits, offering not just discovery potential but also the opportunity to characterise new physics with unprecedented precision.



\begin{figure}[pt]
\centering
\includegraphics[width=0.49\columnwidth]{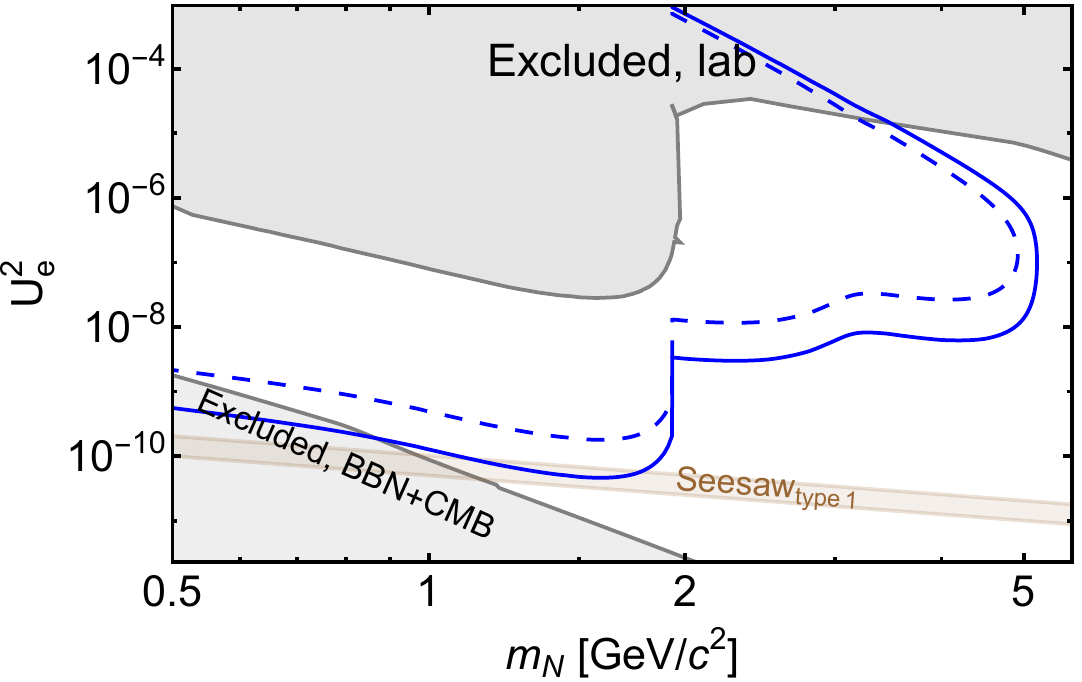}~\includegraphics[width=0.49\columnwidth]{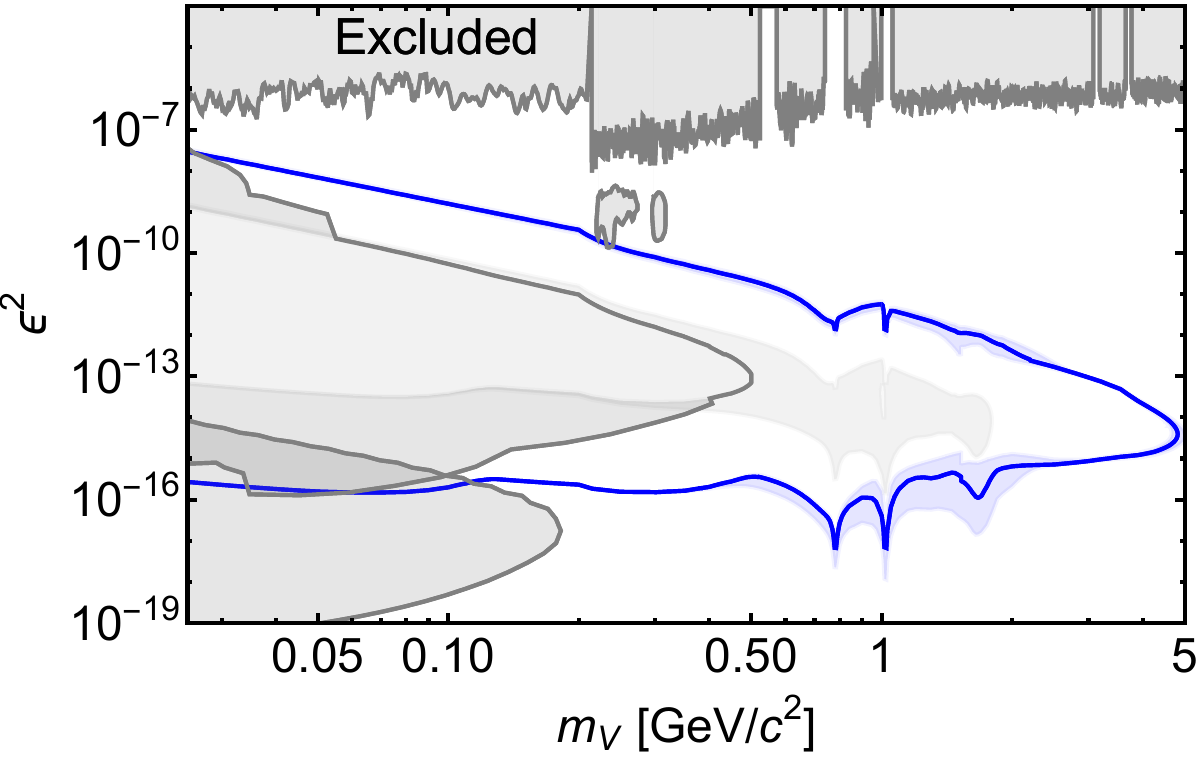}\\\includegraphics[width=0.49\columnwidth]{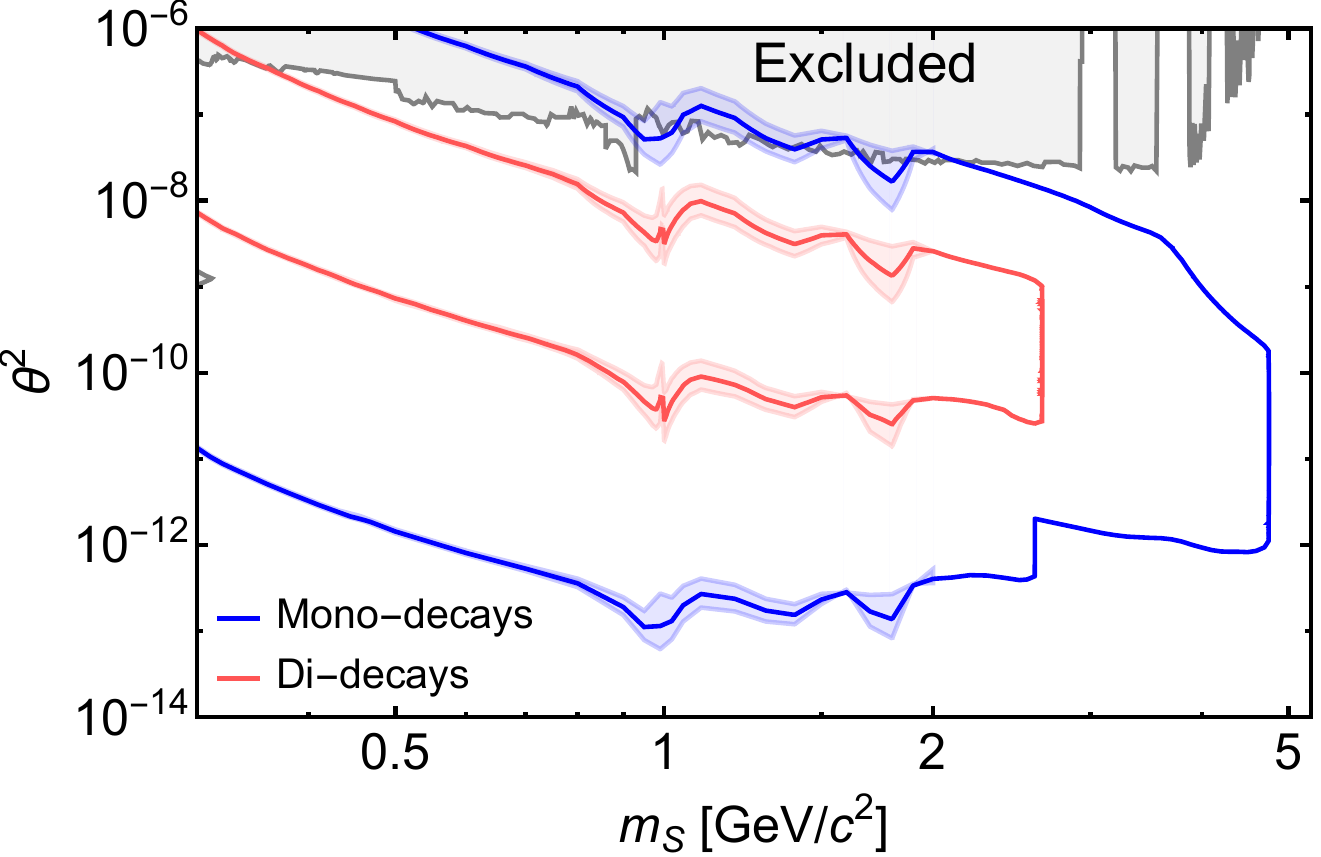}~\includegraphics[width=0.49\columnwidth]{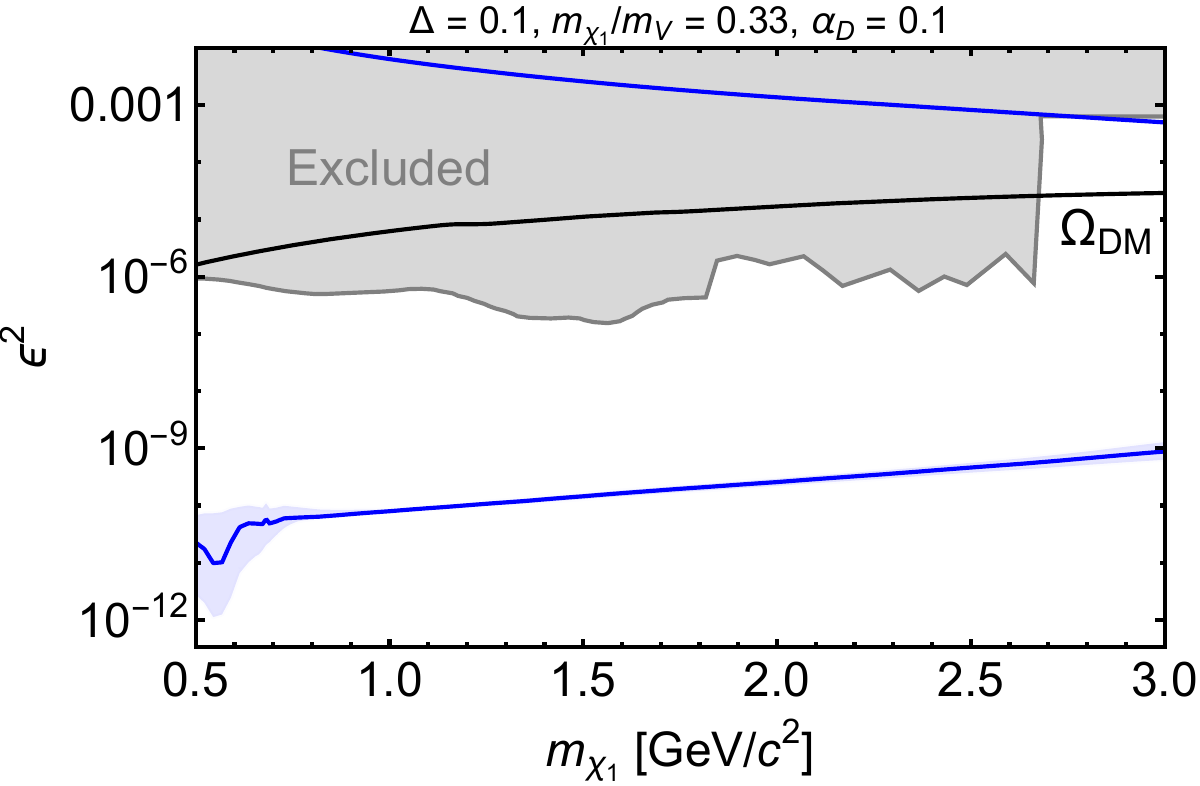}
\caption{Sensitivity of SHiP to various new physics particles: Heavy Neutral Leptons or HNLs (top, left), dark photons (top, right), Higgs-like scalars $S$ with the trilinear coupling to the Higgs boson fixed by $\text{Br}(h\to SS) = 0.01$ (bottom, left), and inelastic dark matter coupled via dark photons (bottom, right). The solid blue lines show the sensitivities per 15-year running time of SHiP. For HNLs, the dashed blue line corresponds to the sensitivity per one-year nominal running time as expected in Run 4. In the case of other models, the relation between the 1-year and 15-year sensitivities is qualitatively the same. For Higgs-like scalars, the red line shows the sensitivity to events with two decaying scalars per event~\cite{DallaValleGarcia:2025aeq}. The light bands around the lines denote the effect of theoretical uncertainty in the production and decays of FIPs on their sensitivity~\cite{Blackstone:2024ouf,Kyselov:2024dmi}; for dark photons, the light-gray region indicates the impact of the uncertainty on the existing constraints.}
\label{fig:FIP_sensitivity}
\end{figure}


Fig.~\ref{fig:FIP_sensitivity} illustrates the projected sensitivity of SHiP for a variety of new physics models, often used in literature as benchmarks. 
Given its robust experimental design and the long-term operation plan, the limits set by SHiP will remain a reference for decades.

Beyond setting new constraints, the redundancy of SHiP’s background rejection systems and its optimized detector layout ensure that, in the event of a discovery, the experiment will not only detect new physics but also precisely measure the properties of any new particles. 
The ability to reconstruct decay vertices, measure lifetimes, and determine coupling strengths will allow for a detailed study of any new phenomena observed. In particular, SHiP can reconstruct the masses of FIPs and distinguish between different theoretical models by analyzing their production and decay characteristics. For instance:
\begin{itemize}
    \item[--] Reconstructing various decay modes of FIPs enables determination of the FIP's spin and the structure of the operators that couple the FIP to the SM~\cite{Mikulenko:2023iqq}. 
    \item[--] HNLs are capable of explaining neutrino oscillations and baryon asymmetry of the Universe. At SHiP, these would manifest themselves via a position-dependent excess of lepton number violating decays and particular patterns of HNL couplings to neutrinos. Observing around 1000 events with HNLs at SHiP may be enough to answer how they are involved in the resolution of those BSM problems~\cite{Mikulenko:2023iqq,Tastet:2019nqj}.
    Furthermore, in the mass range around 1 GeV/c$^2$, SHiP may cover practically the whole parameter space where HNLs is capable of explaining neutrino oscillations (i.e., the HNL couplings above the so-called ``seesaw limit''.)
    \item[--] SHiP's detector for scattering signatures and a possible extension of SHiP with a liquid argon detector may search for unique signatures and identify classes of FIPs~\cite{Ferrillo:2023hhg}: double-scattering signatures with soft recoil (smoking gun signature for millicharged particles), the combined signature with scattering and subsequent decay (typical signatures for HNLs coupled via dipole operator and inelastic dark matter).
\end{itemize}

\subsection{Neutrino physics}

Proton interactions on the BDF target produce a very intense neutrino flux of all three flavours. The use of a long and high-density target is ideal to enrich the neutrino yield of tau neutrinos, since most of the light long-lived mesons are forced to interact before they decay, thus enhancing the fraction of neutrinos from the prompt charmed and beauty hadron decays. This feature greatly enhances the tau and the electron neutrino components over the muon neutrinos, allowing unique studies of their interactions with matter. More than 90\% of the electron neutrinos interacting in the neutrino detector target are expected to originate from charmed hadron decays. Therefore, systematic uncertainties from the neutrino flux cancel out in the $\nu_e/\nu_{\tau}$ ratio. This opens the door to lepton flavour universality studies with neutrino interactions with unprecedented accuracy.   
 
The collected datasets will enable a rich programme of neutrino physics measurements, providing valuable SM tests as well as BSM searches, and serving as an essential benchmark for future neutrino experiments. 

Table~\ref{tab:neutrinoyield} shows the expected neutrino yield in 15 years of operation, assuming a 3-tonne neutrino detector. 
A key objective of SHiP is the first direct observation of tau anti-neutrinos, which have yet to be directly detected experimentally. Until now, only nine tau neutrino candidate events were reported by the DONUT experiment~\cite{DONUT_nutau} without distinguishing between neutrinos and anti-neutrinos, and ten muon-to-tau neutrino oscillations were observed by the Opera experiment~\cite{OPERA_nutau}. The only leptonic decay observed by OPERA shows negative charge as expected from a tau neutrino interaction. 

By exploiting the high-intensity proton beam at the SPS and using a magnetised detector configuration, SHiP will measure both tau neutrino and anti-neutrino interactions with high statistics. This will turn into the only precise study of tau-neutrino cross-sections, significantly improving our understanding of neutrino-nucleon interactions in the 10\,--\,100\,GeV energy range.

Beyond total cross-sections, SHiP will perform detailed studies of deep inelastic scattering (DIS) structure functions, specifically measuring for the first time the F$_4$ and F$_5$ structure functions that are unique to tau neutrino interactions~\cite{Alekhin:2015byh}. These measurements are crucial for refining theoretical models of neutrino-nucleus interactions, which play a fundamental role in current and future long-baseline neutrino oscillation experiments.

Another essential contribution of SHiP to neutrino physics is the study of neutrino-induced charm production~\cite{DeLellis:2004ovi,Alekhin:2015byh}. This process is particularly sensitive to the strange-quark content of nucleons and provides an independent probe of the partonic structure of the nucleon. Precise measurements of charm production in neutrino interactions will also allow a direct determination of the Cabibbo-Kobayashi-Maskawa (CKM) matrix element $|V_{cd}|$. Currently, the uncertainty on $|V_{cd}|$ is dominated by theoretical uncertainties in charm production modeling. The high-statistics sample of neutrino-induced charm events at SHiP will offer a novel approach to determine $|V_{cd}|$ with improved precision, complementing measurements from meson decays and lattice QCD calculations.

Finally, SHiP will also explore BSM physics closely related to neutrinos. Examples include an excess of $\tau$ neutrinos, where hypothetical new physics mediators are efficiently produced in proton collisions and then decay into $\nu_{\tau}$~\cite{Kling:2020iar}, and exotic neutrino up-scattering events, producing FIPs that either decay after traveling a macroscopic distance or escape the detection~\cite{Ferrillo:2023hhg}.

\begin{table}[htp]
	\begin{center}
		\begin{tabular}{c | c c  }
			& CC DIS  &  Charm CC DIS  \\
			\hline
			$N_{\nu_e}$                 & $2.0 \times 10^{6}$ & $1.2 \times 10^{5}$ \\
			$N_{\nu_\mu}$               & $5.8 \times 10^{6}$ & $2.8 \times 10^{5}$ \\
			$N_{\nu_\tau}$              & $5.9 \times 10^{4}$ & $3.2 \times 10^{3}$ \\
			$N_{\overline{\nu}_e}$      & $4.0 \times 10^{5}$ & $2.1 \times 10^{4}$ \\
			$N_{\overline{\nu}_\mu}$    & $1.3 \times 10^{6}$ & $5.0 \times 10^{4}$ \\
			$N_{\overline{\nu}_\tau}$   & $4.3 \times 10^{4}$ & $2.5 \times 10^{3}$ \\
			\hline
		\end{tabular}
	\end{center}
	\caption{Neutrino charged-current deep inelastic interactions (left) and those with a charmed hadron in the final state (right), assuming a mass of 3\,tonnes and 15 years of operation. }
        \label{tab:neutrinoyield}
	\end{table}

{\bf SHiP's dedicated detector for neutrinos and scattering signatures 
will provide not only a crucial test of the SM in the neutrino sector but also ensure that SHiP delivers impactful physics results independently of potential discoveries in the hidden sector. The guaranteed nature of these neutrino measurements strengthens SHiP’s role in the global neutrino-physics landscape.}


\subsection{Search for Light Dark Matter with the neutrino detector}

The lack of detection of Weakly Interacting Massive Particles (WIMPs), which were viable dark matter candidates in the mass range $m\gg 1\text{ GeV}$, has led to increased interest in models of light dark matter (LDM). Such particles could interact via new mediators, such as dark photons, axion-like particles, fundamental scalars, etc. Depending on the interaction type, there are various signatures with LDM that can be probed at SHiP: decays, scatterings of electrons or nucleons, as well as the combination of the latter, such as ``double-bang'' events~\cite{Ferrillo:2023hhg}. We report here the case of LDM scattering off electrons, which may be searched for using the SHiP's neutrino detector.

The final state is characterised by just a single recoil electron. The background source consists of neutrino interactions with an electron in the final state and one or more unidentified particles. The background suppression relies on a high-precision tracking performance to identify any other charged particle produced at the vertex, as well as photons produced by $\pi^0$ decays. Background from deep inelastic scattering, the leading process in the 10\,--\,100 GeV energy range is negligible thanks to the spatial resolution of the high-precision vertex detector. The residual background mainly consists of neutrino elastic scattering.  
For a detector mass of about 3\,tonnes and $6 \times 10^{20}$ protons on target, after topological and kinematical cuts are applied, SHiP has sensitivity to direct detection down to the expected relic density of dark matter in the mass range from 10 MeV/c$^2$ up to $\sim 2m_{\pi}$ in the most commonly used dark photon-LDM benchmark model~\cite{LDM_sensitivity}.

\section{Beamline and target facility}
\label{sec:facility}


At the SPS, the optimal experimental conditions for BDF/SHiP are obtained using a proton beam energy of $400\gev$ combined with short, one-second slow-extracted spills to exploit fast-pulsed cycles to deliver a high yield of protons on target (PoT). While the SPS has routinely delivered a total of $>5\times 10^{19}$ protons per year in the past, the experimental programmes that have been running since 2013 have only been exploiting approximately $1-1.5\times 10^{19}$ protons per year. CERN currently has no experimental facility that is compatible with the maximum available beam power of the SPS. {\bf The ongoing consolidation of the North Area and upgrades of the CERN injector complex 
has motivated a comprehensive upgrade of the existing ECN3 experimental facility to a state-of-the-art high-intensity experimental facility, capable of fully exploiting the capacity of the SPS in parallel to the operation of the HL-LHC and CERN's other fixed target programmes.} The upgrade of the facility and the construction of BDF is carried out under the auspices of the HI-ECN3 project at CERN.

Several detailed investigations of the proton sharing to the different facilities across the CERN accelerator complex have been carried out~\cite{BDF_YELLOWBOOK, LLOstudy, SHiP_ECN3}. With an average of 200 days of operation per year and an average of 5000 spills per day, the studies confirm the possibility of delivering $10^6$ spills of $4\times 10^{13}$ protons in order to annually deliver a total of $4\times 10^{19}$ PoT per year to BDF/SHiP. 
The working point of $4\times 10^{19}$ PoT per year for BDF/SHiP ensures $1.25\times 10^{19}$ protons to the North Area whilst operating the other SPS beam facilities, and $0.85\times 10^{19}$ protons in case a month is dedicated to ions. It has been verified that there is no technical limitation in the accelerator complex, or in the facility design, to continue operation at $4\times 10^{19}$ PoT per year for 15 years, guaranteeing $6\times 10^{20}$ PoT for BDF/SHiP.

The ECN3 experimental hall is located together with the TCC8 target area at the end of the 750\,m long transfer tunnel downstream of TCC2, which is the target area servicing the other North Area experimental halls EHN1 and EHN2 located on the surface. 
Both ECN3 and TCC8 are located entirely below the natural ground level with a 8\,m thick additional layer of earth added on top. 
No additional beam line is needed to locate BDF in ECN3. The proton beam will be transported from the SPS to ECN3 via the existing primary transfer lines (TT20 and P42). The proton spills for BDF/SHiP will be taken through the beam splitter system and past the primary production targets in TCC2 on dedicated cycles without splitting in order to maintain a high transmission efficiency and reduce beam loss to a minimum for the high intensity beam. 
The current T4 target system for the H6 and H8 beamlines in EHN1 will be upgraded to integrate a vacuum system through the shielding downstream of the T4 target to minimise interactions of the high intensity beam with material on the beam line and to reduce beam loss in the shallow downstream transfer tunnels. As part of the upgrade, the wobbling system used to select the secondary particle type and momentum for the H6/H8 beamlines will include a pulsed wobbling magnet to switch protons towards ECN3 so that SHiP can operate in parallel with the other North Area users in EHN1 and EHN2, by distributing the protons on different cycles.

A set of orthogonally aligned dipole magnets located on the beamline 100\,m upstream of SHiP's production target in TCC8 will produce a circular beam sweep on the target's front face in order to dilute the energy deposited by the beam in the target. To maximise the inelastic proton collisions and at the same time provide the cleanest possible background environment by suppressing decays of pions and kaons to muons and neutrinos, SHiP's proton target should be long and made from a combination of materials with the highest possible atomic mass and atomic number. The baseline target design is composed of water-cooled blocks of titanium-zirconium-doped molybdenum alloy~(TZM), cladded by a tantalum-alloy, in the core of the proton shower, followed by blocks of tantalum-cladded pure tungsten. In total, it corresponds to 12 nuclear interaction lengths. Recent developments improving on this design propose a target made completely from tungsten, cooled by helium. Beam tests with a prototype tungsten target is foreseen for early 2026. {\bf The BDF/SHiP target has to fully absorb the 400\,GeV 2.6\,MJ/pulse every 7.2~seconds, corresponding to roughly 350\,kW of average beam power. It puts the target system in a category that has technological synergies with neutron spallation sources worldwide. Collaboration with existing and future user facilities, such as ISIS/STFC in the UK and ESS in Sweden, are underway}.

The target system is integrated into a free-standing shielded complex in TCC8. The first layer of shielding, aiming at absorbing the showers of secondary particles, consists of 400\,cm of cast iron around the target vessel. The target assembly and this proximity shielding is confined in a low vacuum (10$^{-3}$\,mbar) tank in order to reduce air activation and reduce radiation-accelerated corrosion due to ozone production and various other nitric acid compounds. A second assembly of cast iron and concrete blocks encapsulates the vacuum tank. Immediately downstream of the target, the proximity shield incorporates a copper/tungsten plug to further reduce background over the shortest possible length. The downstream hadron stopper and shielding are thick enough to ensure that the radiation environment in the experimental hall allows non-radiation-tolerant electronics. The hadron stopper section is also magnetised to provide the first section of sweeping power for SHiP's muon shield. 

The nTOF Collaboration in collaboration with the HI-ECN3 project have put forward a proposal to implement an activation station close to the SHiP proton target to profit from the wide-spectrum ultra-high neutron flux. The setup can be operated in synergy with the operation of SHiP without disturbing the main physics programme. It will allow unique measurements of key nuclear reactions for the first time, of urgent need in the fields of nuclear astrophysics, nuclear energy and nuclear medicine~\cite{NTOF_PROPOSAL}.

The HI-ECN3 project is currently in the technical design phase with the delivery of the Technical Design Report for the intensity upgrade and BDF expected mid-2026. The project is working in close synergy with the North Area Consolidation (NA-CONS) project to upgrade the required accelerator infrastructure as part of Phase 1 of the consolidation programme to be implemented during Long Shutdown 3 in 2026 - 2029. This is followed by the construction work of the BDF in TCC8 during the first part of Run 4 in parallel to resumed operation of the North Area experiments, aiming for completion of the BDF ahead of first beam on target in 2031.

The cost of upgrading and implementing BDF in ECN3 for SHiP is approximately 100\,MCHF lower than the ECN4 proposal presented in the 2020 ESPP proposal~\cite{ESPP_BDF2022}. The saving was made by removing the need for major civil engineering works and by re-using existing infrastructure as well as profiting from the ongoing investment in the consolidation of the North Area. Alongside the NA-CONS project, the upgrade of ECN3 will secure the long-term future of the SPS Fixed Target physics programme at CERN.

\begin{figure}[!t]
  \centering
  \includegraphics[width=\textwidth]{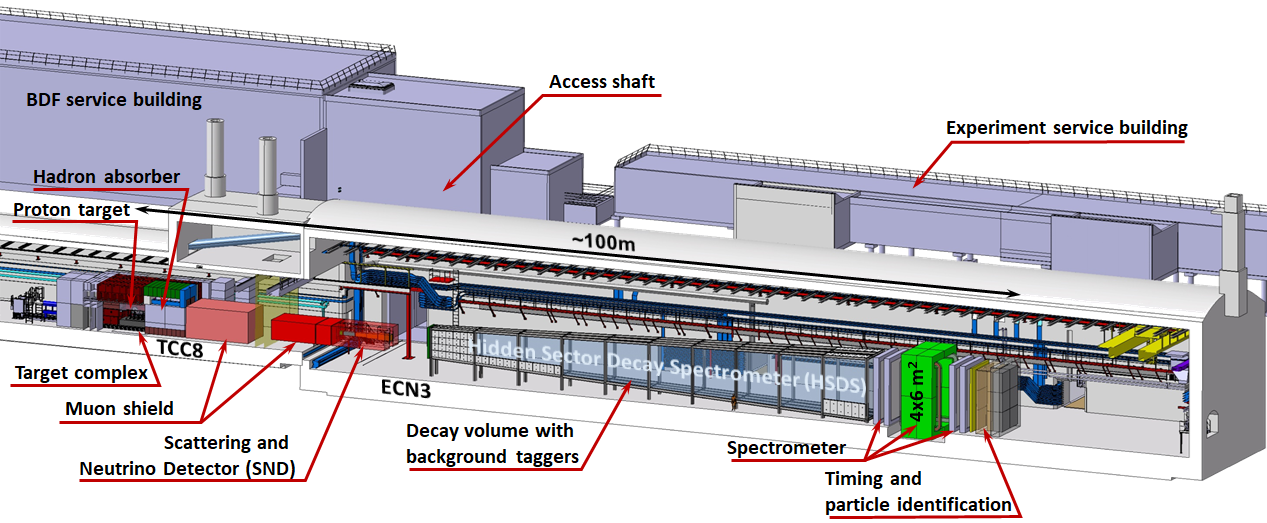}
  \caption{\small Overview of BDF/SHiP in ECN3.}
  \label{fig:ship_overview}  
\end{figure}

\section{SHiP detectors}
\label{detector}


The SHiP experiment is shown in Fig.~\ref{fig:ship_overview}. 
The particles of interest to SHiP are long-lived and can have decay products with large transverse momenta, necessitating a detector which is both long in the beam direction and large in the direction transverse to the beam.


The first component of the experimental setup just downstream of the target complex is the muon shield, built to sweep out the $\sim 20$\,GHz of muons above 10\,GeV produced every SPS spill, leaving only a residual rate of less than $100$\,kHz of muons in the detector acceptance. The muon shield consists of the magnetised hadron stopper and a series of free-standing magnets arranged along $\sim 25$\,m along the beam axis, with transverse dimensions of 1 to 3 metres. An integral field of $\sim 30$\,Tm is required to sweep the highest-momenta muons out of the detector acceptance. Mid-momentum muons deflected in the upstream part of the shield encounter the return field which bends such muons back towards the detector. To solve this problem, the downstream magnets of the muon shield are energized with an opposite polarity. 
This way, the downstream magnets' return field deflects back towards the outside the mid-momentum muons, while it continues to gradually bend out the high-momentum muons. 
Superconducting technology with a 5\,T field is currently under investigation for the upstream section, as it would permit a shorter,
more compact muon shield. 

\begin{figure}[!t]
  \vspace*{-4mm}
  \centering
  \includegraphics[width=\textwidth]{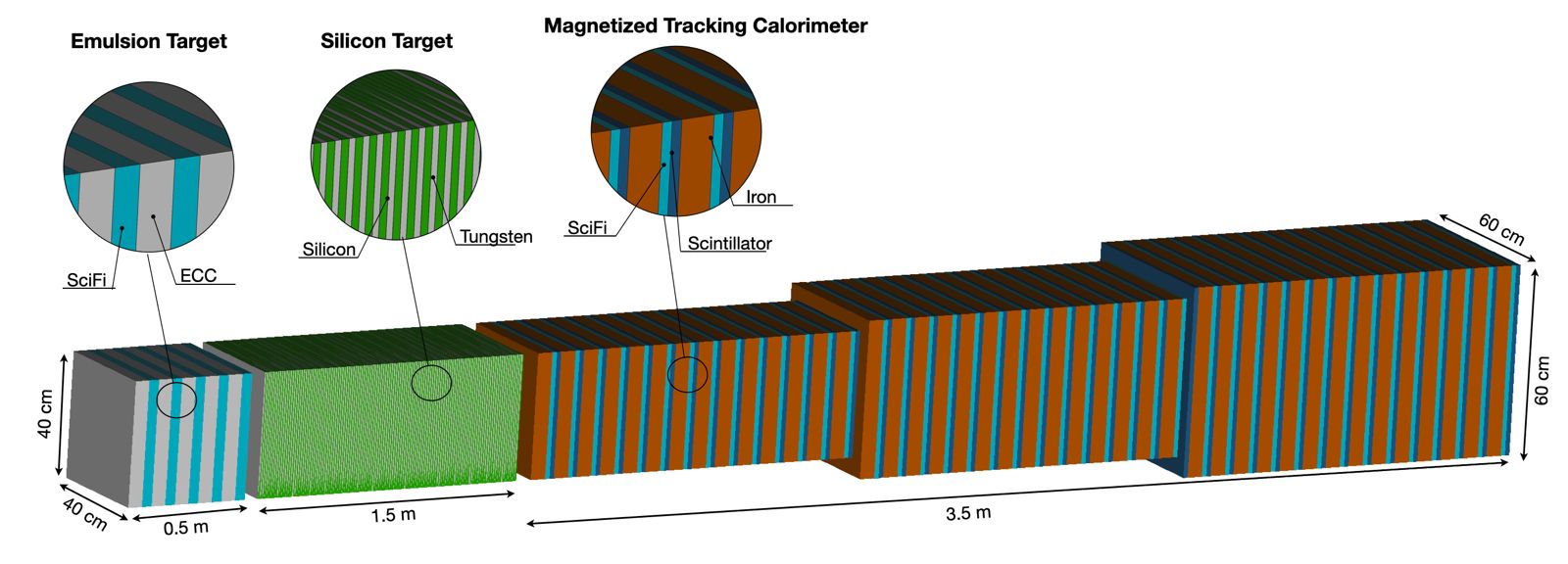}
 \vspace*{-5mm} 
  \caption{\small Schematic drawing of the SND detector.}
  \label{fig:SNDSHiP}  
\end{figure}

The SHiP scattering and neutrino detector, SND, is positioned inside the last part of the muon shield. As shown in Fig.~\ref{fig:SNDSHiP}, it consists of a sandwich of
tungsten plates and silicon detectors, which provide precise vertex reconstruction in tau neutrino events and reconstruction of the electromagnetic showers in electron neutrino events, and a magnetised tracking calorimeter (MTC). Upstream of this setup, an emulsion cloud chamber (ECC) made of tungsten plates interleaved with emulsion films will complement the high-precision tracking device, if the muon rate will be tolerable. The MTC is made of a magnetised iron absorbers, which uses the return field of the central yoke of the muon shield, and active planes of the scintillating-fibres (SciFi) and scintillating tiles. The SciFi planes provide the reconstruction of neutrino interaction vertices and can measure the momenta of charged tracks as well as missing transverse momentum carried away by neutrinos from decays of tau leptons. The scintillating tiles are used to measure the energy of hadron showers. The MTC is designed to reconstruct all neutrino flavours, including tau neutrinos, using kinematical methods. The silicon/tungsten vertex detector can reconstruct tau neutrino events on an event-by-event basis using the MTC for measuring the momenta and energies of muons and hadrons. 

A lightweight gas bag filled with helium at 1\,atm pressure and held in place by a structure of aluminum frames will be used as the experiments search volume for FIP decays. Residual backgrounds will be dominated by muons and the products of neutrino interactions in the material. This necessitates the use of background tagger systems surrounding the decay volume to detect suspicious activity. The taggers will be built from multi-gap resistive plate chambers or scintillating fibres upstream, and from liquid scintillator detector modules mounted onto all sides of the frame of the gas bag and readout by 
wavelength-shifting optical modules. 

A spectrometer magnet with an aperture of 4$\times$6\,m$^2$ and a field integral of 0.6-0.8\,Tm gives SHiP the required momentum resolution for charged decay products. This field will be achieved with a superconducting magnet with a peak field of about 0.15\,T~\cite{Spectrometer_magnet}. The geometry of the tracker requires optimisation with the helium-filled decay volume to ensure the best overall angle determination, which depends on the magnetic field integral and the positioning of the tracking stations. The tracking will be provided by four stations of drift detectors composed of ultralight 4m-long straws. The stations are designed to each be just 0.5\% of a radiation length. 

An integrated particle identification system includes an electromagnetic calorimeter (ECAL) to reconstruct electrons and photons, and a compact hadron sampling calorimeter with a low sampling ratio to provide high-efficiency muon/hadron discrimination in a wide momentum range. Using high-precision layers located at specific shower depths, the ECAL is also capable of measuring the photon direction with a sufficient accuracy to reconstruct two-photon final states of FIPs such as $\mathrm{ALP}\rightarrow\gamma\gamma$.

\section{Collaboration and funding}
\label{status}


SHiP, with its unique physics potential, emerges as a central part of CERN's strategy to ensure a broad, diverse scientific programme, complementary to colliders and exploiting the injectors. CERN is fully committed to the high-intensity upgrade of the ECN3 experimental facility and to the construction of BDF for SHiP. Following the approval of the technical design phase, the collaboration is actively working to secure funding from national and international funding agencies. Given the agreed-upon timeline of the TDRs and the construction, institutions from several countries have already submitted applications or are in the process of applying. In particular, applications have been submitted or are being prepared in Italy, Switzerland, and the UK, with other countries expected to follow soon.

Among the primary responsibilities for the initial configuration of SHiP, the muon shield is a critical component required to ensure the low-background environment necessary for feebly interacting particle searches. The development of the muon shield involves contributions from institutes in Germany, Italy, Switzerland and the UK. The decay volume, veto system, tracker and spectrometer, including timing, involve institutes in Germany, Italy, and Switzerland, together with CERN. 

The SHiP collaboration is also making significant efforts to expand its institutional participation. Currently, the collaboration consists of 36 institutes in 18 countries, with 9 additional associate institutes, and CERN and JINR. Several new institutes have  expressed interest in joining the experiment. 


{\bf The coming months will be crucial in consolidating institutional commitments and aligning funding strategies with national agency priorities. SHiP’s role as a key component of CERN’s intensity-frontier programme strengthens the case for broad international support, ensuring that the experiment can move forward in line with the planned timeline for construction and commissioning. }



\begin{thebibliography}{9}

\bibitem{Bonivento:2013jag}
W. Bonivento et al., {\it Proposal to Search for Heavy Neutral Leptons at the SPS}, CERN-SPSC-2013-024, SPSC-EOI-010, \href{https://arxiv.org/abs/1310.1762}{[arXiv:1310.1762]}.

\bibitem{Anelli:2015pba}
SHiP Collaboration, {\it A facility to Search for Hidden Particles (SHiP) at the CERN SPS}, Technical Proposal, CERN-SPSC-2015-016, SPSC-P-350,  \href{https://arxiv.org/abs/1504.04956}{[arXiv:1504.04956]}.

\bibitem{Alekhin:2015byh}
S. Alekhin et al., {\it A facility to Search for Hidden Particles at the CERN SPS: the SHiP physics case}, Rept. Prog. Phys. 79 (2016), no. 12 124201, \href{https://arxiv.org/abs/1811.00930}{[arXiv:1811.00930]}.

\bibitem{SHiP_CDS}
SHiP collaboration, {\it SHiP Experiment - Comprehensive Design Study report}, Tech.
Rep. CERN-SPSC-2019-049, SPSC-SR-263, CERN, Geneva (Dec, 2019).

\bibitem{SHiP_detector}
[12] SHiP collaboration, {\it The SHiP experiment at the proposed CERN SPS Beam Dump
Facility}, Eur. Phys. J. C 82 (2022) 486, \href{https://arxiv.org/abs/1811.00930}{[arXiv:1811.00930]}.

\bibitem{SHiP_ECN3}
SHiP collaboration, {\it BDF/SHiP at the ECN3 high-intensity beam facility - Technical Proposal}, Tech. Rep. CERN-SPSC-2023-033, SPSC-P-369, CERN, Geneva (2023).

\bibitem{SHiP:2018yqc}
SHiP collaboration, {\it The experimental facility for the Search for Hidden Particles at
the CERN SPS}, JINST 14 (2019) P03025, \href{https://arxiv.org/abs/1811.00930}{[arXiv:1811.00930]}.

\bibitem{BDF_YELLOWBOOK}
C. Ahdida et al., {\it SPS Beam Dump Facility - Comprehensive Design Study}, CERN
Yellow Reports: Monographs, CERN, Geneva (Dec, 2019),
10.23731/CYRM-2020-002,\href{https://arxiv.org/abs/1912.06356}{[arXiv:1912.06356]}.

\bibitem{LLOstudy}
O. Aberle et al., {\it Study of
alternative locations for the SPS Beam Dump Facility}, Tech. Rep.
CERN-SPSC-2022-009, SPSC-SR-305, CERN, Geneva (2022), \href{https://arxiv.org/abs/2204.03549}{[arXiv:2204.03549]}.

\bibitem{Asaka:2005an}
T. Asaka, S. Blanchet, M. Shaposhnikov, {\it The $\nu$MSM, Dark Matter and Neutrino Masses}, Phys. Lett. B631 (2005), p151-156, \href{https://arxiv.org/abs/0503065}{[arXiv:hep-ph/0503065]}.

\bibitem{Mikulenko:2023iqq} 
O. Mikulenko, K. Bondarenko, A. Boyarsky, O. Ruchayskiy, {\it Unveiling new physics with discoveries at Intensity Frontier}, \href{https://arxiv.org/abs/2312.05163}{[arXiv:2312.05163]}.

\bibitem{Tastet:2019nqj} 
J.L. Tastet, I. Timiryasov, {\it Dirac vs. Majorana HNLs (and their oscillations) at SHiP}, JHEP 04 (2020) 005, \href{https://arxiv.org/abs/1912.05520}{[arXiv:1912.05520]}.

\bibitem{Ferrillo:2023hhg} 
M. Ferrillo, M. Ovchynnikov, F. Resnati, A. De Roeck, {\it Improving the potential of BDF\@SPS to search for new physics with liquid argon time projection chambers}, JHEP 02 (2024) 196, \href{https://arxiv.org/abs/2312.14868}{[arXiv:2312.14868]}.

\bibitem{DallaValleGarcia:2025aeq}

G. Dalla Valle Garcia, M. Ovchynnikov, {\it Di-decay signature of new physics particles at intensity frontier experiments}, \href{https://arxiv.org/abs/2503.01760}{[arXiv:2503.01760]}.

\bibitem{Blackstone:2024ouf}

P. J. Blackstone et al., {\it Hadronic Decays of a Higgs-mixed Scalar}, \href{https://arxiv.org/abs/2407.13587}{[arXiv:2407.13587]}.

\bibitem{Kyselov:2024dmi}

Y. Kyselov, M. Ovchynnikov, {\it Searches for long-lived dark photons at proton accelerator experiments}, Phys.Rev.D 111 (2025) 1, 015030, \href{https://arxiv.org/abs/2409.11096}{[arXiv:2409.11096]}.

\bibitem{DeLellis:2004ovi} 
G. De Lellis et al., {\it Charm physics with neutrinos}, Physics Reports 399 (2004) 227. 

\bibitem{DONUT_nutau}
K.K. et al., {\it Final tau-neutrino results from the DONuT experiment}, Phys. Rev. D - Particles, Fields, Gravitation and Cosmology 78, pp. 1-20 (2008).

\bibitem{OPERA_nutau}
OPERA collaboration, {\it Final Results of the OPERA Experiment on $\nu_{\tau}$ Appearance
in the CNGS Neutrino Beam}, Phys. Rev. Lett. 120 (2018) 211801 \href{https://arxiv.org/abs/1804.04912}{[arXiv:1804.04912]}.

\bibitem{Kling:2020iar}
F. Kling, {\it Probing light gauge bosons in tau neutrino experiments}, Phys. Rev. D 102 (2020) no. 1, 015007, \href{https://arxiv.org/abs/2005.03594}{[arXiv:2005.03594]}.

\bibitem{LDM_sensitivity}
SHiP collaboration, {\it Sensitivity of the SHiP experiment to light dark matter}, JHEP
04 (2021) 199, \href{https://arxiv.org/abs/2010.11057}{[arXiv:2010.11057]}.

\bibitem{NTOF_PROPOSAL}
nTOF Collaboration, {\it Neutron Activation Station at the SPS Beam Dump Facility (BDF) - Letter of Intent}, Tech. Rep. CERN-SPSC-2024-027 / SPSC-EOI-023, CERN, Geneva, 2024.

\bibitem{ESPP_BDF2022}
C. Ahdida et al., {\it SPS Beam Dump Facility - Comprehensive overview}", submitted to EPPSU, 2018.

\bibitem{Spectrometer_magnet}
A. Devred et al., {\it Proof-of-Principle of an Energy-Efficient, Iron-Dominated Electromagnet for Physics Experiments}, IEEE Transactions on Applied Superconductivity, Vol. 34, No. 5, August 2024.

\end{thebibliography}




\onecolumn

\noindent
\textbf{SHiP Collaboration}

\noindent
R.~Albanese$^{16,d,f}$,
K.~Albrecht$^{7}$,
J.~Alt$^{8}$,
A.~Alexandrov$^{16,d}$,
F.~Alessio$^{29}$,
S.~Aoki$^{19}$,
D.~Aritunov$^{10}$,
N.~Azorskiy$^{40}$, 
L.~Baudin$^{29}$,
V.~Bautin$^{40}$
% ...

\vspace{0.5cm}

\noindent

\vspace*{1cm}


{\footnotesize \it

\noindent
% ...
$^{7}$Humboldt-Universit\"{a}t zu Berlin, Berlin, Germany\\
$^{8}$Physikalisches Institut, Universit\"{a}t Freiburg, Freiburg, Germany\\
% ...
$^{10}$Forschungszentrum J\"{u}lich GmbH (KFA),  J\"{u}lich , Germany\\
% ...
$^{16}$Sezione INFN di Napoli, Napoli, Italy\\
% ...
$^{19}$Kobe University, Kobe, Japan\\
% ...
$^{29}$European Organization for Nuclear Research (CERN), Geneva, Switzerland\\
% ...
$^{40}$Joint Institute for Nuclear Research, Dubna, Russia \\
% ...
$^{d}$Associated to Gyeongsang National University, Jinju, Korea\\
% ...
$^{f}$Universit\`{a} della Basilicata$^{j}$, Potenza, Italy\\
% ...
$^{j}$Associated to Universit\`{a} di Napoli ``Federico II``, Napoli, Italy\\
% ...
}


We acknowledge contributions from 
A.~Anokhina, 
E.~Atkin, 
A.~Bagulya, 
A.Y.~Berdnikov, 
Y.A.~Berdnikov, 
M.~Chernyavskiy, 
V.~Dmitrenko, 
A.~Etenko, 
O.~Fedin, 
K.~Filippov, 
G.~Gavrilov, 
V.~Giuliaeva, 
V.~Golovtsov, 
D.~Golubkov, 
S.~Gorbunov, 
M.~Gorshenkov, 
V.~Grachev, 
V.~Grichine,
N.~Gruzinskii, 
Yu.~Guz, 
D.~Karpenkov, 
M.~Khabibullin, 
E.~Khalikov, 
A.~Khotyantsev, 
V.~Kim, 
N.~Konovalova, 
I.~Korol'ko, 
Y.~Kudenko, 
P.~Kurbatov, 
V.~Kurochka, 
E.~Kuznetsova,  
V.~Maleev, 
A.~Malinin, 
A.~Mefodev, 
O.~Mineev, 
S.~Nasybulin,
B.~Obinyakov, 
N.~Okateva, 
A.~Petrov, 
D.~Podgrudkov, 
M.~Prokudin,  
T.~Roganova, 
V.~Samsonov, 
E.S.~Savchenko, 
A.~Shakin, 
P.~Shatalov, 
T.~Shchedrina, 
V.~Shevchenko, 
A.~Shustov, 
M.~Skorokhvatov, 
S.~Smirnov, 
N.~Starkov, 
P.~Teterin, 
S.~Than~Naing, 
S.~Ulin,  
A.~Ustyuzhanin, 
Z.~Uteshev, 
L.~Uvarov,
K.~Vlasik, 
A.~Volkov, 
R.~Voronkov, 
A.~Zelenov.

\end{document}